\documentclass{article}
\usepackage[preprint]{spconf,amsmath,graphicx}
\usepackage{booktabs}
\usepackage{amssymb}
\usepackage{pifont}
\usepackage[hidelinks]{hyperref}
\usepackage{siunitx}
\usepackage{multirow}

\title{How much to Dereverberate? Low-Latency Single-Channel Speech Enhancement in Distant Microphone Scenarios}
\name{Satvik Venkatesh, Philip Coleman, Arthur Benilov, Simon Brown, Selim Sheta, Frederic Roskam}
\address{L-Acoustics, 67 Southwood Lane, London N65EG.}
\copyrightnotice{\begin{tabular}[t]{@{}l@{}}© 2025 IEEE. Personal use of this material is permitted. Permission from IEEE must be obtained for all other uses, in any current or\\future media, including reprinting/republishing this material for advertising or promotional purposes, creating new collective works,\\for resale or redistribution to servers or lists, or reuse of any copyrighted component of this work in other works.\end{tabular}}

\begin{document}
\ninept
\maketitle
\begin{abstract}
Dereverberation is an important sub-task of Speech Enhancement (SE) to improve the signal's intelligibility and quality. However, it remains challenging because the reverberation is highly correlated with the signal. Furthermore, the single-channel SE literature has predominantly focused on rooms with short reverb times (typically under 1~s), smaller rooms (under volumes of $10^3 m^3$) and relatively short distances (up to 2 meters). In this paper, we explore real-time low-latency single-channel SE under distant microphone scenarios, such as 5 to 10 meters, and focus on conference rooms and theatres, with larger room dimensions and reverberation times. Such a setup is useful for applications such as lecture demonstrations, drama, and to enhance stage acoustics. First, we show that single-channel SE in such challenging scenarios is feasible. Second, we investigate the relationship between room volume and reverberation time, and demonstrate its importance when randomly simulating room impulse responses. Lastly, we show that for dereverberation with short decay times, preserving early reflections before decaying the transfer function of the room improves overall signal quality.
\end{abstract}
\section{Introduction}
\label{sec:intro}

Speech enhancement (SE) or speech restoration tries to improve the intelligibility and quality of speech contaminated by additive noise \cite{defossez2019music, liu2022voicefixer}, reverberation~\cite{luo2018real, maciejewski2020whamr}, clipping, and low sampling rates~\cite{kuleshov2017audio}. The topic has been addressed by various machine learning challenges such as Deep Noise Suppression (DNS)~\cite{dubey2022icassp, dubey2023icassp}, REVERB~\cite{kinoshita2016summary}, Computation Hearing in Multisource Environments (CHiME)~\cite{cornell2023chime}, and spatialized DNS~\cite{pandey2022tparn, lee2023deft}. SE algorithms that work in real-time with low-latency are useful for applications such as online meetings~\cite{defossez2020real}, hearing aids~\cite{van2009speech}, or as a pre-processing step for speech recognition and separation~\cite{maciejewski2020whamr, vincent2017analysis}. 

Among the signal degradations mentioned above, removing reverberation (or dereverberation) presents a distinct challenge. Unlike noise degradations which are typically additive in nature, reverberation is convolutive and consequently the undesired reverberation is correlated with the desired signal.
SE studies have approached reverberation in multiple ways. The DNS Challenge in 2021 mentioned that the SE model should input noisy reverberant speech and output clean reverberant speech~\cite{reddy2021icassp}. Dereverberation is optional and not a requirement. Other state-of-the-art studies such as the FullSubNet+~\cite{chen2022fullsubnet} investigated SE under reverberant and non-reverberant conditions, but did not explicitly dereverberate the signal. 

Despite this, late reverberation, in particular, is known to degrade the intelligibility of speech~\cite{hu2014effects}. Research in Deep Neural Networks has made great progress in SE under reverberant conditions~\cite{zhou2023speech}. 
The literature for single-channel SE with low-latency has predominantly focused on relatively small rooms such as offices and homes; and close-mic scenarios (typically under 2 meters), where the direct-to-reverberant ratio (DRR) is relatively high. In this paper, we explore this task in larger conference rooms, theatres, and auditoria, which are expected to have greater reverberant energy. Furthermore, increasing the distance between the microphone and the source to 5 or 10 meters significantly reduces the DRR. While array processing solutions such as beamformers have been shown to work under distant microphone scenarios, the single-channel distant microphone SE problem has not been explored in the literature to our knowledge. In this paper, we demonstrate the feasibility of low-latency SE in large rooms at large talker-to-mic distances. We also demonstrate how the quality can be improved by generating training data with the volume and $T_{60}$ constrained in a way that matches the typical physical acoustics of the rooms of interest. 

Further, SE algorithms sometimes aim to preserve some early reflections, as they are highly correlated with the direct sound and support  intelligibility~\cite{zhou2023speech}. For speech, the early reflections are generally considered to be those arriving within 50~ms of the direct sound. However, as \cite{zhou2023speech} pointed out, abruptly truncating these early reflections sounds unnatural because such a room does not exist in reality. Therefore, studies have explored the idea of decaying the transfer function of the room~\cite{braun2021towards, zhou2023speech, schroter2023deepfilternet}. In this method, the target signal to train the neural network is generated by convolving the clean speech with a shorter version of the original room impulse response, for instance, $T^{max}_{60} = 300~ms$~\cite{braun2021towards}. We therefore also investigate, for our distant mic SE scenario, how much residual reverberation to leave in the target signal, and in what scenarios the early reflections are useful. 
Finally, taking into account the insights we have gained in dataset generation, we demonstrate the efficacy of our training pipeline for distant microphone scenarios by comparing with state-of-the-art models.

\section{Proposed Method}
\subsection{Volume-based $T_{60}$ Sampling}
Most studies for dereverberation use synthetic RIRs generated through the image source method (ISM), or a hybrid model that combines ISM and diffuse reverberation~\cite{scheibler2018pyroomacoustics, maciejewski2020whamr, diaz2021gpurir}. These RIR simulators are fed random values within specified ranges. For example, room dimensions in the range of (3, 3, 2.5) m and (10, 10, 5) m. In addition, $T_{60}$ is randomly sampled within a range such as 0.1 to 0.8~s~\cite{zhang2022multi}. However, these studies do not consider the relationship between room volume and $T_{60}$. This is less crucial in small rooms and small Reverberation Times (RTs). When simulating large rooms with large RTs, we are likely to create unrealistic RIRs using this naive approach. For example, consider the room with dimensions (40, 40, 20) m. The volume of this room is $3.2 \times 10^4~m^3$ and it is unrealistic to have a $T_{60}$ as low as 0.1~s. Furthermore, in a small room such as (3, 3, 2.5)~m, we cannot have a $T_{60}$ of 1.8~s, as it will be a very nasty sounding room. Such situations need to be avoided when generating synthetic RIRs, otherwise the neural network will learn data that is non-representative of real-world scenarios. 

The $T_{60}$ of a room depends on various factors such as volume, surface area of walls, materials used, and furniture. In architecture, there are different guidelines for different types of rooms. Some examples of $T_{60}$ for different room types with different volumes can be found in table \ref{table:t60_room_types}. These values were obtained from \cite{harris1957handbook}.

\begin{figure}[t] 
\centering
\footnotesize
\includegraphics[width=8.5cm]{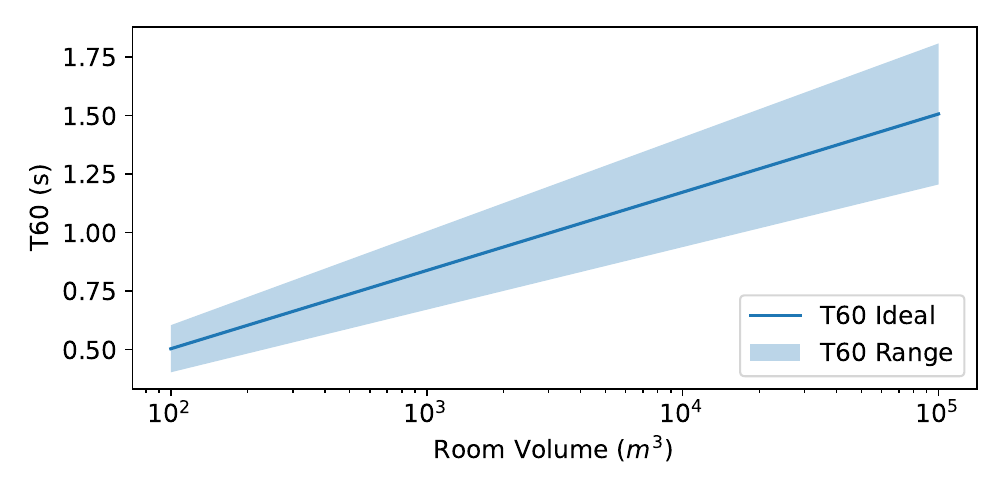}
\vspace{-.2cm}
\caption{The proposed relationship between room volume and $T_{60}$ for conference rooms, curve-fitted with values from ~\cite{harris1957handbook, cirillo2007acoustics, othman2012influence}.}
\vspace{-.2cm}
\label{fig:rt60_range}
\end{figure}

\begin{table}[t]
\centering
\vspace{-.22cm}
\caption{$T_{60}$ depending on room type and volume~\cite{harris1957handbook, cirillo2007acoustics, othman2012influence}.}
\label{table:t60_room_types}
\footnotesize
\begin{tabular}{@{}lll@{}}
\toprule
Room Type         & Volume ($m^3$)         & $T_{60}$ (s)        \\ \midrule
Radio Studio      & 100, 500, 2000      & 0.4, 0.75, 1.2 \\
Catholic church   & 500, 1000, 5000     & 1.3, 1.5, 1.8  \\
Speech auditorium & 200, $10^3$, $10^4$ & 0.7, 0.8, 1.0  \\ 
Conference room & 200, $10^3$, $10^4$ & 0.6, 0.84, 1.17  \\
\bottomrule
\end{tabular}
\end{table}
We investigated the relationships in the literature between reverb time and volume for conference rooms~\cite{harris1957handbook, cirillo2007acoustics, othman2012influence} and used curve-fitting to summarise them into equation~\ref{eq:t60_vol}.
\begin{equation}
    \label{eq:t60_vol}
    T_{60} = a \cdot ln(V) - b
\end{equation}
where $V$ is the volume of the room, $a=0.145$ and $b=0.165$; considering a variation of $\pm 20\%$. Figure \ref{fig:rt60_range} plots this equation.

\subsection{Windowing Room Impulse Responses}
\label{sec:boost_clarity}
Braun et al. \cite{braun2021towards} shaped room impulse responses to a maximum decay of $T^{max}_{60} = 300~ms$. A constant window $w(n)$ as defined in equation \ref{eq:window} is multiplied by the room impulse responses to obtain the target for the neural network to learn. 
\begin{equation}
\label{eq:window}
    w(n) = 
    \begin{cases}
      1 & \text{for $n \leq N_{1}$} \\
      10^{-q(n - N_{1})} & \text{for $n > N_{1}$}
    \end{cases}
\end{equation}
where $q = 3/(T^{max}_{60} \cdot f_{s})$ and $N_{1}$ is the direct sound. This method has also been adopted by other SE studies such as NSNet~\cite{braun2021towards} and DeepFilterNet~\cite{schroter2022deepfilternet2, schroter2023deepfilternet} with different values of $T^{max}_{60}$. Zhou et al. \cite{zhou2023speech} proposed reverb-time shortening, where they naturally decay the room impulse response to a target $T_{60}$, instead of a constant window function. Here, the decay rate of the window is adaptive and depends on the $T_{60}$ of the original impulse response. In this case, $q = \{3/(T^{max}_{60} \cdot f_{s})\} - \{3/(T_{60} \cdot f_{s})\}$. However, the literature has not yet compared the use of constant versus varying windows for dereverberation. In this paper, we use constant windows as they are easier to control and examine the impact of different $T^{max}_{60}$ values such as 150, 300, and 500~ms under distant microphone scenarios.

\begin{figure}[t] 
\centering
\includegraphics[width=8.5cm]{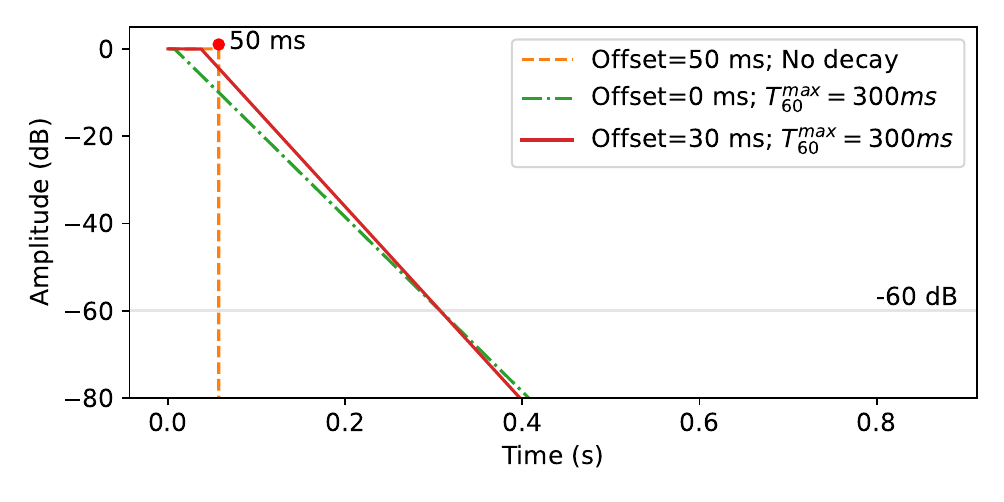}
\vspace{-.2cm}
\caption{Gain curves applied for reverb suppression.}
\label{fig:decay-offset}
\vspace{-.2cm}
\end{figure}

Furthermore, as shown in figure~\ref{fig:decay-offset}, we investigate the benefit of having an offset before naturally decaying the room impulse response. Therefore, offset is added to $N_1$ in equation \ref{eq:window} and is subtracted from $T^{max}_{60}$ to ensure the curves intersect at -60~dB. The motivation behind this idea is to simultaneously preserve intelligibility and naturally decay the room impulse response. The DeepFilterNet3~\cite{schroter2023deepfilternet} model uses an offset of 5~ms and target $T^{max}_{60}$ of 500~ms. In this paper, we explore other combinations of offsets and shorter $T^{max}_{60}$ values in distant microphone scenarios.





\section{Experimental Setup}

\subsection{Dataset}
We used the clean speech, noise, and RIR dataset provided by the DNS challenge 2022~\cite{dubey2022icassp}. The clean speech is a corpora of various datasets such as VCTK~\cite{yamagishi2019vctk}, PTDB~\cite{pirker2011pitch}, and read speech from Wall street journal~\cite{dubey2022icassp}. To manage compute storage constraints, we trained only on the English dataset, similar to \cite{schroter2022deepfilternet2}. To generate RIRs for distant microphone scenarios, we used FRA-RIR~\cite{luo2022fra} and gpuRIR~\cite{diaz2021gpurir}. 
The Signal-to-Noise Ratio (SNR) was randomised between (-5, 40)~dB during training and (5, 40)~dB during testing. For the test set, we collected 37 RIRs for distant microphone scenarios from \cite{coleman2020s3a}, \cite{murphy2010openair}, and in-house recordings to generate 100 audio examples. The same test set was used for all sections in the paper. 






\subsection{Nework Architectures}
We explore speech enhancement under similar constraints as the DNS challenge, that is to work in real-time with a latency of 20~ms~\cite{dubey2023icassp} or 40~ms~\cite{dubey2022icassp} at 48~kHz sampling rate. We adapted training pipeline and data augmentation methods from DeepFilterNet~(DFN), whose source-code is openly available~\cite{schroter2022deepfilternet2}. We explored training two neural network architectures in this study, (1) the state-of-the-art DFN3 model which has a latency of 40~ms (2) Hybrid Spectrogram Time-domain Audio Separation Network-Small (HSTN), proposed by \cite{venkatesh2024real} originally for real-time low-latency music source separation, which has a latency of 20~ms. We benchmarked the models on the Voicebank-Demand dataset to ensure our training pipeline is comparable to the state-of-the-art architectures such as FRCRN~\cite{zhao2022frcrn} and FullSubNet+~\cite{chen2022fullsubnet}. As the Voicebank-Demand is not the main focus of this study, these results are provided as supplementary material.







\subsection{Parameters for Dereverberation}
We consider four scenarios: \\
    \noindent \textbf{Close Mic. Small Room}: The distance between the microphone is in the range of 0.1 to 0.5~m. The minimum and maximum room dimensions are (3, 3, 2.5)~m and (10, 10, 5)~m respectively. \\
    \noindent \textbf{Close Mic. Large Room}: The distance between the microphone is in the range of 0.1 to 1~m. The minimum and maximum room dimensions are (3, 3, 2.5)~m and (40, 40, 20)~m respectively. \\
    \noindent \textbf{Medium Mic. Small Room}: The mic. distance is in the range of 0.1 to 2~m. The room parameters are the same as the first scenario. \\
    \noindent \textbf{Far Mic. Large Room}: The mic. distance is in the range of 0.2 to 10~m. The room parameters are the same as the second scenario.

We performed Analysis of Variance (ANOVA) to investigtate the effects of early reflections and $T^{max}_{60}$. We considered early offsets of 0, 5, 30, 50, and 80 ms; and $T^{max}_{60}$ values of No Decay (N.D.), 150, 300, and 500 ms. We performed post-hoc t-tests with Bonferroni correction for pairwise comparisons.



\section{Results}
To evaluate the SE models, we adopt Perceptual Evaluation of Speech Quality (PESQ), Short-Time Objective Intelligibility (STOI), and Scale-Invariant Signal-to-Distortion Ratio (SI-SDR). We also present the Deep Noise Suppression Mean Opinion Score (DNSMOS), which is a neural network that predicts the perceptual evaluation score for three factors --- signal quality (SIG), background noise suppression (BAK), and overall quality (OVRL)~\cite{reddy2021dnsmos}.

\subsection{Volume-based $T_{60}$ Sampling}
In this subsection, we trained only on the VCTK clean speech examples to manage computational costs. We use synthetic RIRs during training and real-world RIRs for testing. For the close mic. setting, we only use RIRs provided by the DNS challenge. For the far mic. without volume-based sampling, we randomise room dimensions between (3, 3, 2.5)~m to (40, 40, 20)~m and $T_{60}$ values between 0.1 to 1.8~s. For the configuration with volume-based sampling, it is the same room dimensions, but with the $T_{60}$ defined by equation \ref{eq:t60_vol} and $\pm 20\%$ variation. For the test set, we collected RIRs for distant microphone scenarios from \cite{coleman2020s3a}, \cite{murphy2010openair}, and in-house recordings to generate 100 audio examples of 10~s each. 

In table \ref{table:result_vol}, we compare noisy, close mic., and far-mic. without and with volume-based $T_{60}$ sampling. Interestingly, the difference between close mic. and far mic. without volume-based sampling is negligible. This conveys that the neural network does not learn new meaningful information from RIRs without volume-based sampling. However, with volume-based $T_{60}$ sampling, the PESQ increases from 2.08 to 2.17, STOI from 85.9\% to 87.6\%, and OVRL DNSMOS from 2.64 to 2.69. This conveys the importance of considering the volume of the room when sampling $T_{60}$ values.

\begin{table}[t]
\addtolength{\tabcolsep}{-0.2em}
\centering
\footnotesize
\vspace{-.22cm}
\caption{Comparing close mic., far mic. without volume-based $T_{60}$ sampling, and with volume-based $T_{60}$ sampling.}
\label{table:result_vol}
\begin{tabular}{@{}ccccccc@{}}
\toprule
\multirow{2}{*}{Model} & \multirow{2}{*}{PESQ} & STOI & SI-SDR & MOS   & MOS   & MOS    \\
                       &                       & (\%) & & (SIG) & (BAK) & (OVL) \\ \midrule
Noisy                  & 1.49                   & 77.8 & 3.61  & 1.93 & 2.01  & 1.62   \\
Close Mic              & 2.08                  & 85.9 & 6.72   & 2.92 & \textbf{3.94}  & 2.64   \\
Far Mic w/o vol        & 2.09                  & 86.1 & 6.70  & 2.91 & 3.93  & 2.62   \\
Far Mic w vol          & \textbf{2.17}                  & \textbf{87.6} & \textbf{7.50} & \textbf{2.98}  & 3.92  & \textbf{2.69}    \\ \bottomrule
\end{tabular}
\end{table}

\subsection{Window Design for Dereverberation}
\label{sec:params_dereverberation}
\begin{figure*}[tb] 
\centering
\includegraphics[width=17.2cm]
{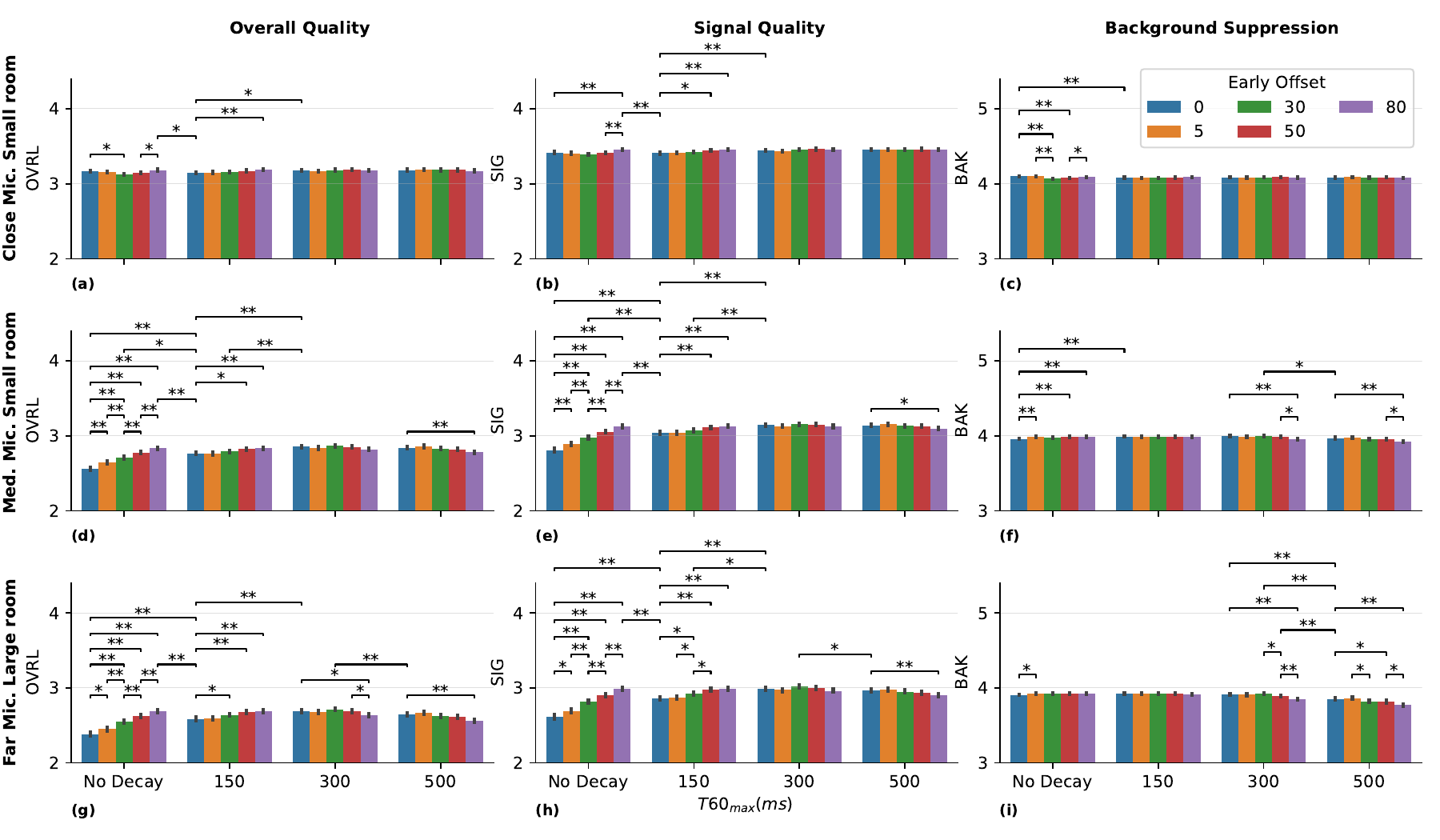}
\vspace{-.4cm}
\caption{DNSMOS OVRL, SIG, and BAK scores for the model trained with different dereverberation parameters tested under close and distant microphone scenarios in small and large rooms. The asterisks indicate the level of significance (*: $p < 0.05$, **: $p < 0.001$).}
\label{fig:results_main_significance}
\vspace{-.2cm}
\end{figure*}

In this subsection, we again train only on VCTK clean speech as we are testing many hypotheses. Moreover, we synthesised the test set impulse responses using gpuRIR~\cite{diaz2021gpurir} so that we could precisely control the room dimensions and distance from the microphone. We consider DNSMOS~\cite{reddy2021dnsmos} instead of intrusive metrics such as PESQ and STOI because there is no appropriate ground truth.

ANOVA indicated that both offset and $T^{max}_{60}$ have a significant effect on the DNSMOS scores. Figure~\ref{fig:results_main_significance} shows the OVRL, SIG, and BAK for close/distant microphone scenarios in small and large rooms. A more comprehensive plot with numerical values is available as supplementary material here\footnote{\label{footnote:supplementary}\href{https://l-acoustics.github.io/icassp2025.github.io/}{https://l-acoustics.github.io/icassp2025.github.io/}}.

\subsubsection{Close Microphone}
In the small room, the direct sound (0, N.D.) obtains a high OVRL performance. None of the settings are significantly better than the direct sound. Interestingly, preserving early reflections of 30 ms (30, N.D.) is significantly worse than the direct sound due to poorer BAK performance. 
Therefore, for small distances under 0.5~m, the network does not require early reflections or to naturally decay the impulse response. We can simply predict the direct sound.

The results for the large room are not shown in figure~\ref{fig:results_main_significance}, but plotted in the supplementary material\textsuperscript{\ref{footnote:supplementary}}. We observe very similar patterns as the small room close microphone setting, which conveys that hyperparameters for dereverberation depend largely on the distance from the microphone and less on the room dimensions.

In both close microphone scenarios, we observe that $T^{max}_{60}$ has a negligible effect on performance. Therefore, there is no evident benefit in preserving reverberant energy for close microphone scenarios.

\subsubsection{Distant Microphone}
In the small room, we can observe significant improvements in OVRL and SIG score when preserving early reflections for $T^{max}_{60}=N.D$. The OVRL scores are 2.56, 2.64, 2.71, 2.78, and 2.83 for 0, 5, 30, 50, and 80~ms respectively. This demonstrates the importance of early reflections in distant microphone scenarios. 

We also observe a significant improvement in performance when the $T^{max}_{60}$ is increased from N.D. to 150 and 300 ms. The OVRL scores are 2.56, 2.77, and 2.86 for (0, N.D.), (0, 150), and (0, 300) respectively. However, we do not see improvements when increasing the $T^{max}_{60}$ to 500~ms. Furthermore, preserving too much reverberant energy degrades BAK performance. This can be observed in pairwise comparisons of ((0, 300) \& (80, 300)) and ((0, 500) \& (80, 500)). The reason for this could be that the network is trying to predict the late diffused reverberation, which is essentially noise. In addition, the background noise gets mixed with the late energy, which makes it challenging for the network to differentiate between the two. In the small room, (0, 300) and (30, 300) obtain the highest OVRL score of 2.86.

In the far microphone large room setting, similar to the small room, we observe significant effects for early reflections. For N.D., the OVRL scores are 2.38, 2.45, 2.56, 2.62, and 2.69 for 0, 5, 30, 50, and 80~ms respectively. Furthermore, in the large room, the degradation in BAK performance around $T^{max}_{60}=500 ms$ is more severe than the small room. This is probably due to large rooms having more late reverberant energy than small rooms.

Interestingly, preserving early energy can compensate for the smaller $T^{max}_{60}$. For example, (80, N.D.) is significantly better than (0, 150); the difference between (50, 150), (80, 150) and (0, 300) is small and not significant. Therefore, in cases where a short $T^{max}_{60}$ such as 150~ms is preferred, having an offset of at least 30~ms is beneficial. Moreover, for higher $T^{max}_{60}$ such as 300~ms, having an offset of greater than 0 is less beneficial. In addition, 300~ms is the $T_{60}$ of most studio rooms and hence, is generally an acceptable threshold for residual reverberation~\cite{braun2020data}. (30, 300) obtains the highest OVRL score of 3.02, compared to 2.99 for (0, 300). 
Although the difference between (0, 300) and (30, 300) is negligible, we select (0, 300) because it is more common in the literature and has also been used to train the NSNet2~\cite{braun2021towards}.









\subsection{Comparison with State-of-the-art}
In table \ref{table:final}, we present HSTN and DFN3 models trained under distant microphone scenarios (DFN3-d.m. and HSTN-d.m.) with a dereverberation target of (0, 300). Compared to the original DFN3 model, the OVRL DNSMOS score improves from 2.77 to 3.04. This shows the robustness of our training pipeline for distant microphone SE. The original DFN3 was trained with a target of (5, 500), different from (0, 300). Thus, we do not present the intrusive metrics--- PESQ, STOI, and SI-SDR, as these values are lower and give a wrong representation of the model's performance. The HSTN model obtains an OVRL DNSMOS score of 2.87, which is still higher than the other models in the literature, and with a lower latency of 20~ms. Audio examples from different models are available\textsuperscript{\ref{footnote:supplementary}}.

\begin{table}[h]
\addtolength{\tabcolsep}{-0.2em}
\centering
\footnotesize
\vspace{-.2cm}
\caption{Comparing our distant microphone (d.m.) adaptations of HS-TasNet (HSTN) and DeepFilterNet3~(DFN3) with other state-of-the-art SE models. ${\dagger}$ indicates from current work. }
\label{table:final}
\begin{tabular}{@{}cccccccc@{}}
\toprule
\multirow{2}{*}{Model} & \multirow{2}{*}{PESQ} & STOI & SI-SDR & MOS   & MOS   & MOS  & Lat.  \\
                       &                       & (\%) & & (SIG) & (BAK) & (OVL) & (ms)\\ \midrule
Noisy                  & 1.49                   & 77.8 & 3.61 & 1.93 & 2.01  & 1.62 & -  \\
NSNet2~\cite{braun2020data}              & 2.17                  & 86.8 & 7.19 & 2.93 & 3.92  & 2.64 & 20  \\
FSN+~\cite{chen2022fullsubnet}              & -                  & - & - & 2.48 & 2.90  & 2.09 & 32  \\
DFN3~\cite{schroter2023deepfilternet}              & -                  & - & - & 3.10 & 3.90  & 2.77 & 40  \\
DFN3-d.m.$^{\dagger}$           & \textbf{2.59}    & \textbf{90.9} & \textbf{9.49} & \textbf{3.32} & \textbf{4.05}  & \textbf{3.04} & 40 \\
HSTN-d.m.$^{\dagger}$  & 2.36   & 89.6 & 8.80 & 3.15 & 4.01  & 2.87 & 20 \\ \bottomrule
\end{tabular}
\vspace{-.4cm}
\end{table}

\section{Conclusion}
In this paper, we investigated real-time low-latency SE under distant microphone scenarios. We demonstrated that SE under such challenging scenarios is feasible and obtained state-of-the-art performance for this task. When simulating RIRs, it was helpful to consider the volume of the simulated room. This helps us generate more realistic RIRs, improve stability of the training pipeline, and effectively improve SE performance. 

Later, the results in section \ref{sec:params_dereverberation} showed that the room size was less important when tuning the parameters for dereverberation. Instead, the most important factor was the distance between the source and the microphone. As the distance from the microphone increases, the early reflections become more correlated with the direct sound, which makes it more challenging for the network. Hence, at larger distances, preserving early reflections helps the network. We found that having an offset of at least 30~ms is beneficial for a short $T^{max}_{60}$ such as 150~ms, but had no significant effect for 300~ms. In future work, listening tests may help us better understand the trade-offs between different values of offset and $T^{max}_{60}$. 

Distant microphone SE is useful for applications such as lecture demonstrations, drama, and to enhance stage acoustics. In future work, domain adaptation methods that address the domain shift from close to distant microphone scenarios could improve performance. A recently introduced signal improvement challenge~\cite{cutler2024icassp} that focuses on addressing distortions such as colouration, discontinuities, and reverberation is relevant to this study too. 




\vfill\pagebreak

\footnotesize
\bibliographystyle{IEEE}
\bibliography{refs}

\end{document}